\documentclass[a4paper,12pt]{article}
\usepackage{amssymb}
\usepackage{cite}
\usepackage[english]{babel}

\newcommand{\dual}{{}^*}

\newcommand{\dd}{{\rm d}}

\newcommand{\re}{\mathrm{Re}}
\newcommand{\im}{\mathrm{Im}}

\newcommand{\eq}{\begin{equation}}
\newcommand{\feq}{\end{equation}}
\newcommand{\eqn}{\begin{eqnarray}}
\newcommand{\feqn}{\end{eqnarray}}
\newcommand{\arr}{\begin{eqnarray*}}
\newcommand{\farr}{\end{eqnarray*}}

\newcommand{\RR}{{\cal R}}

\newcommand{\F}{{\cal F}}
\newcommand{\G}{{\cal G}}
\newcommand{\p}{\partial}

\font\mybb=msbm10 at 12pt
\def\bb#1{\hbox{\mybb#1}}

\def\bR {\bb{R}}

\newcommand{\HH}{{\mathbb{H}}}

\newcommand{\sder}{\mathcal{D}} % Superderivative operator

\def\al{\alpha}

\def\ep{\epsilon}

\def\si{\sigma}

\def\Ga{\Gamma}

\begin{document}
\begin{titlepage}
\begin{flushright}
IFUM-809-FT\\
hep-th/0411153
\end{flushright}
\vspace{.3cm}
\begin{center}
\renewcommand{\thefootnote}{\fnsymbol{footnote}}
{\Large \bf On Supersymmetric Solutions of $D = 4$ Gauged Supergravity\footnote{Talk given at the ``Workshop on Dynamics and Thermodynamics of Black Holes and Naked Singularities'', Politecnico di Milano, 13-15 May 2004}}
\vskip 25mm
{\large \bf {Marco M.~Caldarelli\footnote{\tt marco.caldarelli@mi.infn.it}}}\\
\renewcommand{\thefootnote}{\arabic{footnote}}
\setcounter{footnote}{0}
\vskip 10mm
{\small
Dipartimento di Fisica dell'Universit\`a di Milano\\
and\\
INFN, Sezione di Milano,\\
Via Celoria 16, I-20133 Milano.\\
}
\end{center}
\vspace{2cm}
%\begin{center}
%\end{center}
\begin{abstract}
\noindent
We review the classification of supersymmetric solutions to minimal gauged supergravity in four dimensions. After a short introduction to the main features of the theory, we explain how to obtain all its solutions admitting a Killing spinor. Then, we analyze the rich mathematical structure behind them and present the supersymmetric field configurations. Among them, we find supersymmetric black holes, quarter and half BPS traveling waves, kink solutions, and supersymmetric Kundt and Robinson-Trautman solutions. Finally, we generalize the classification to include external sources, and show a particular solution describing a supersymmetric G\"odel-type universe.
\end{abstract}

\end{titlepage}

%%%%%%%%%%%%%%%%%%%%%%%%%%%%%%%%%%%%%%%%%%%%%%%%%%%%%%%%%%%%%%%%%%%%%%%%%%%%%%

\section{Introduction}
\label{intro}

Recently, significant steps have been made towards a classification of supersymmetric solutions of supergravity theories. Apart from the interest in understanting the mathematical structure lying behind these BPS configurations, such investigations are important due to the role supersymmetry has played in the developments in string theory. Having at hand a systematic approach to construct BPS solutions avoids the use of special ans\"atze and has lead to the discovery of many new supergravity backgrounds.

After the seminal work by Tod \cite{Tod:pm}, who wrote down all metrics admitting supercovariantly constant spinors in $N=2$, $D=4$ ungauged supergravity, progress in this direction has been made mainly during the last years using the mathematical concept of G-structures \cite{Gauntlett:2002sc}.
This formalism has been applied successfully to several supergravity theories \cite{sugra,Gauntlett:2003fk,Caldarelli:2003pb}.
New interesting solutions have been obtained using these techniques, among which there are a maximally supersymmetric G\"odel universe in five dimensions \cite{Gauntlett:2003fk}, supersymmetric AdS black holes in five dimensions \cite{Gutowski}, D1-D5-P black hole microstates \cite{d1d5} in the spirit of the proposal by Mathur et al. \cite{Mathur:2003hj},
and half BPS excitations of AdS$\times{\mathcal S}$ configurations described by free fermions in the dual field theories \cite{Lin:2004nb}.
Moreover, the results for the simpler, lower-dimensional supergravities can be uplifted to ten or eleven dimensions to describe some sectors of ten- and eleven-dimensional supergravities. 

In this article, we will review the classification of supersymmetric solutions to minimal gauged supergravity in four dimensions \cite{Caldarelli:2003pb,Caldarelli:2003wh,Cacciatori:2004rt}, which extends the early results \cite{Romans:1991nq,Kostelecky:1995ei,Caldarelli:1998hg,Cacciatori:1999rp,Alonso-Alberca:2000cs,Brecher:2000pa}, where only plane waves in AdS and BPS solutions belonging to the general Petrov type D solution of Pleba\'nski and Demia\'nski \cite{Plebanski:gy} were analyzed.

We will begin with a short introduction to minimal gauged supergravity in four dimensions and a presentation of the technique used to find all its supersymmetric solutions \cite{Caldarelli:2003pb}. These fall in two cases, the ``lightlike case'' and the ``timelike case,'' to be defined below. In section \ref{lightlike} we shall present the complete classification of the lightlike solutions, showing that there is an enhancement of the supersymmetry for a subfamily of solutions. Next, in section~\ref{timelikecase}, we shall present the general timelike solution, describe some properties of the set of equations governing it and solve them in some cases, obtaining interesting new solutions \cite{Cacciatori:2004rt}. In section \ref{external} we will extend the theory to include external sources, and build a supersymmetric G\"odel universe\cite{Caldarelli:2003wh}. The final section will be devoted to conclusions.

%%%%%%%%%%%%%%%%%%%%%%%%%%%%%%%%%%%%%%%%%%%%%%%%%%%%%%%%%%%%%%%%%%%%%%%%%%%%%%

\section{Minimal gauged supergravity} \label{sugra}
The gauged version of $N=2$ supergravity in four dimensions was found by Freedman and Das \cite{Freedman:1976aw} and by Fradkin and Vasiliev
\cite{Fradkin:1976xz}. In this theory, the rigid $\mathrm{SO(2)}$ symmetry
rotating the two independent Majorana supersymmetries present in the ungauged
theory is made local by introduction of a minimal gauge coupling between the
photons and the gravitini. Local supersymmetry then requires a negative
cosmological constant and a gravitini mass term.
The theory has four bosonic and four fermionic degrees of freedom;
it describes a graviton $e_\mu^a$, two real gravitini $\psi_\mu^i$ $(i=1,2)$,
and a Maxwell gauge field ${\mathcal A}_{\mu}$. As we said, the latter is
minimally coupled to the gravitini, with coupling constant $\ell^{-1}$.
The bosonic trucation of the lagrangian reads
\begin{equation}
e^{-1}L = -\frac{1}{4}R
            + \frac{1}{4}{\mathcal F}_{\mu\nu}{\mathcal F}^{\mu\nu}
            - \frac{3}{2\ell^2}
            \label{lagrange}
\end{equation}
and describes the Einstein-Maxwell theory with a negative cosmological constant $\Lambda = -3\ell^{-2}$. We are interested in supersymmetric solutions of this model, i.e.~bosonic solutions whose supersymmetry variation vanishes.
A local supersymmetry transformation generated by an infinitesimal Dirac spinor $\epsilon$ produces a variation of the gravitino field equal to $\delta\psi_\mu=\sder_\mu\epsilon$, where $\psi_\mu=\psi_\mu^1+i\psi_\mu^2$ and the supercovariant derivative is defined by
\begin{equation}
\sder_\mu = \nabla_\mu  - \frac i\ell{\mathcal A}_\mu
+ \frac{1}{2\ell}\Ga_\mu 
+ \frac{i}{4}\hat{\mathcal F}_{ab}\Ga^{ab}
                 \Ga_\mu\,. \label{supcovder}
\end{equation}
Hence, a bosonic background is invariant under some local supersymmetry transformations if there is some non-trivial solution to the Killing spinor equation
\begin{equation}
\sder_\mu \epsilon = 0\,. \label{killspinequ}
\end{equation}
This equation implies the following integrability condition,
\begin{eqnarray}
[\sder_{\nu}, \sder_{\mu}]\epsilon &=& \left[\frac 1\ell(\dual{\mathcal F}_{\nu\mu}\Gamma_5
- i{\mathcal F}_{\nu\mu}) + \frac 1{2\ell^2}\Gamma_{\nu\mu} +
\frac 14{{\mathcal R}^{ab}}_{\nu\mu}\Gamma_{ab}\right. \nonumber \\
&& - {\mathcal F}^{\alpha\beta}{\mathcal F}_{\beta [\nu}\Gamma_{\mu]\alpha}
+ \frac 14 {\mathcal F}_{\alpha\beta}{\mathcal F}^{\alpha\beta}\Gamma_{\nu\mu}
- \frac i{\ell}{{\mathcal F}^{\alpha}}_{[\nu}\Gamma_{\mu]\alpha} \nonumber \\
&& \left.-\frac i2 \Gamma_{\alpha\beta[\nu}\nabla_{\mu]}{\mathcal F}^{\alpha\beta}
- i\nabla_{[\nu}{{\mathcal F}_{\mu]}}^{\alpha}\Gamma_{\alpha}\right]\epsilon = 0\,.
\label{intcond}
\end{eqnarray}
For a solution to be maximally supersymmetric, condition (\ref{intcond}) must impose no algebraic constraint on the Killing spinor, and therefore {\em the only maximally supersymmetric geometry is given by AdS$_4$ with vanishing gauge field}. An analogous result in five dimensions was obtained in \cite{Gauntlett:2003fk}.

To find the most general BPS solution we suppose that the geometry admits at least one Killing spinor $\epsilon$, which can be used to construct some bosonic differential forms out of its bilinears. These forms are not all independent, and in fact there are some algebraic relations linking them, derived from the Fierz identities. Particularly important are the two one-forms $V_\mu=i\bar\ep\Ga_\mu\ep$ and $A_\mu=i\bar\ep\Ga_5\Ga_\mu\ep$. A first consequence of the algebraic relations is that $V^2\leq0$, and therefore the vector $V^\mu$ is either timelike or null.
For simplicity, we will name {\em timelike solutions} the solutions which have a Killing spinor yielding a timelike vector $V$ and {\em lightlike solutions} the ones giving a null vector $V$. If the solution preserves just one quarter of the supersymmetry, the timelike or lightlike character of the solution
% are mutually exclusive, and 
produces a rough classification. However, if there is some additional supersymmetry, the solution can fall in both classes, depending on the particular Killing spinor used to define the vector $V$.

Next, the Killing spinor equation $\sder_\mu \epsilon = 0$ imposes a set of differential constraints for these forms. In particular, one finds $\nabla_{(\mu}V_{\nu)}=0$ and $\dd A=0$ that allow the introduction of two preferred coordinates $t$ and $z$, defined by $V=\p_t$ and $A=\dd z$. Then, by cleverly using the algebraic and differential constraints and imposing the equations of motion for the gauge field, one obtains the most general form of the metric and gauge fields preserving some supersymmetry \cite{Caldarelli:2003pb}. In the timelike case, the Einstein equations automatically follow from the integrability conditions for the Killing spinor, whereas in the lightlike case the component along $V\otimes V$ of these equations has to be additionally imposed. Finally, one obtains a last equation requiring that the integrability conditions are satisfied, to ensure the actual existence of the Killing spinor.

In the following we will analyze the solutions that arise, and review the progress made towards a complete classification. We shall begin with the lightlike case, which is simpler and allows for an exhaustive classification. Then, we shall turn to the more complicated timelike case.
%%%%%%%%%%%%%%%%%%%%%%%%%%%%%%%%%%%%%%%%%%%%%%%%%%%
%%%%%%%%%%%%%%%%%%%%%%%%%%%%%%%%%%%%%%%%%%%%%%%%%%%
\section{The lightlike case}

\label{lightlike}

The general lightlike supersymmetric solution is an electrovac traveling wave with metric and electromagnetic field given by \cite{Caldarelli:2003pb}
\begin{eqnarray}
  \dd s^2&=&\frac{\ell^2}{x^2}\left[{\mathcal G}(x,y,u)\,\dd u^2+2\,\dd u\dd v
  +\dd x^2+\dd y^2\right]\,,
\label{lobwave}\\
  {\mathcal F}&=&\dd{\mathcal A} =\varphi'(u)\,\dd u\wedge\dd x\,,\qquad
  {\mathcal A}=\varphi(u)\,\dd x\,.
\end{eqnarray}
Here, the arbitrary function $\varphi'(u)$ defines the profile of the
electromagnetic wave propagating on this metric, while $\G(x,y,u)$ is
any solution of the inhomogeneous Siklos equation \cite{Siklos:1985}\footnote{In this review, $\Delta=\p^2_x+\p^2_y$ denotes the two-dimensional flat laplacian in $(x,y)$ plane.}
\begin{equation}
  \Delta{\mathcal G}-\frac2x\partial_x{\mathcal G}=-\frac{4x^2}{\ell^2}
  \left(\varphi'\right)^2\,.
\label{siklos}\end{equation}
The dependence of $\G(x,y,u)$ on $u$ describes the profile of the
gravitational wave.
The general solution to this equation reads \cite{Siklos:1985}
\begin{equation}
  \G(\zeta,\bar\zeta,u)=\frac12\left(\zeta+\bar\zeta\right)\left(
    \partial f+\bar\partial\bar f\right)-\left(f+\bar f\right)
  -\frac{\left(\varphi'(u)\right)^2}{16\ell^2}\left(\zeta+\bar\zeta\right)^4\,,
  \label{gensol}
\end{equation}
where $f(\zeta,u)$ is an arbitrary holomorphic function in $\zeta=x+iy$.

This family of traveling waves enjoys a large group of coordinate
transformations which preserve the form (\ref{lobwave}) of the line
element. Under the diffeomorphism
$(u,v,x,y)\mapsto(\bar u,\bar v,\bar x,\bar y)$ defined by
\begin{eqnarray}
  \bar u =\chi(u)\,,\qquad \bar x=x\sqrt{\chi'(u)}\,,\qquad
  \bar y=y\sqrt{\chi'(u)}-\psi(u) \,, \phantom{\frac12}\nonumber\\
  \bar v=v-\frac{\chi''(u)}{4\chi'(u)}\left(x^2+y^2\right)
  +\frac{\psi'(u)}{\sqrt{\chi'(u)}}y+\gamma(u)\,,\!\!\!\qquad\qquad\qquad
\label{diff}\end{eqnarray}
where $\chi(u)$, $\psi(u)$ and $\gamma(u)$ are arbitrary functions of
$u$, the metric keeps the same form (\ref{lobwave}) but with
$\bar\G(\bar x,\bar y,\bar u)$ given by \cite{Siklos:1985}
\begin{eqnarray}
  \bar\G(\bar x,\bar y,\bar u)=\frac1{\chi'(u)}\left[
    \G(x,y,u)+\frac12\left\{\chi(u);u\right\}\left(x^2+y^2\right)-2\gamma'(u)
  \right]\nonumber\\
  -\frac{2y}{\sqrt{\chi'(u)}}\left(\frac{\psi'(u)}{\chi'(u)}\right)'
  -\left(\frac{\psi'(u)}{\chi'(u)}\right)^2\,.\qquad\qquad\qquad
\label{Gtrans}\end{eqnarray}
Here, the prime denotes the derivative with respect to $u$, while
\begin{equation}
\left\{\chi(u);u\right\}=\frac{\chi'''(u)}{\chi'(u)}-\frac32\left(\frac{\chi''(u)}{\chi'(u)}\right)^2
\end{equation}
defines the Schwarzian derivative. These are not Killing symmetries,
the metric changes, but solutions are brought into other solutions.
The special diffeomorphisms with $\psi(u)=\gamma(u)=0$ correspond to
reparameterizations of the coordinate $u$; this transformation group
is generated by a central extension of the Virasoro algebra
\cite{Banados:1999tw}.

To obtain a complete classification of these solutions one has to solve the Killing spinor equation and obtain the exact fraction of supersymmetry preserved by them.
This can be done exactly (see \cite{Cacciatori:2004rt}), and surprisingly it turns out that a subclass of these solutions has an enhanced supersymmetry. It is useful to consider separately the cases with and without gauge field excitations.

\subsection{Purely gravitational waves}

This case, with $\varphi'(u)=0$, has been extensively studied in \cite{Siklos:1985}. Two cases are possible, and the vanishing of $\left(\Delta_-\G\right)^2+ 4\left(\G_{,xy}\right)^2$, with $\Delta_-\equiv\p_x^2-\p_y^2$ discriminates them.

\paragraph{The maximally supersymmetric case:}
The condition $\Delta_-\G=\G_{,xy}=0$ is satisfied if and only if the spacetime is AdS \cite{Siklos:1985}.
In this case no condition is imposed on the Killing spinor by the integrability conditions, and one recovers the well-known result that AdS is a maximally supersymmetric spacetime \cite{Breitenlohner:jf}. This is the unique maximally supersymmetric solution of the model under consideration \cite{Caldarelli:2003pb}.

\paragraph{One quarter BPS Lobatchevski waves:} When $\Delta_-\G\neq0$
or $\G_{,xy}\neq0$, the spacetime describes an exact AdS gravitational wave \cite{Podolsky:1997ik}.
The integrability conditions project out half of the components of the Killing spinor, but while solving the Killing spinor equation, an additional condition on the Killing spinor emerges, leaving just one quarter of the supersymmetries, as shown in \cite{Brecher:2000pa}. This is a very simple example that shows that the vanishing of the supercurvature is not a sufficient condition to ensure the
existence of Killing spinors.

\subsection{The electromagnetic case}

Let us now turn on the electromagnetic field and consider the
case where $\varphi'(u)\neq0$.
Generically, these solutions preserve one quarter of the supersymmetries. However, if the solution of the Siklos equation assumes a very particular form, then there is a supersymmetry enhancement. It was shown in \cite{Cacciatori:2004rt} that for
\begin{equation}
  \G(x,y,u)=-\frac1{\ell^2}\varphi'(u)^2x^4+\frac16\xi_3(u)x^3
  +\frac12\xi_2(u)\left(x^2+y^2\right)+\xi_1(u)y+\xi_0(u)
  \label{Ghalf}\end{equation}
where $\xi_0(u)$ and $\xi_1(u)$ are arbitrary functions and
$\xi_2(u)$ and $\xi_3(u)$ are given by
\begin{eqnarray}
  \xi_2(u)&=&\frac89\left(\frac{\varphi''(u)}{\varphi'(u)}\right)^2-\frac23
  \frac{\varphi'''(u)}{\varphi'(u)}-2\alpha^2\ell^2\left[\varphi'(u)\right]^{4/3}\nonumber\\
  \xi_3(u)&=&\displaystyle12\alpha\left[\varphi'(u)\right]^{5/3}\,.
  \label{xi}\end{eqnarray}
{\em the supergravity solution preserves exacly one half of the supersymmetries}. 
Recently, the same phenomenon has been noted in the case of five-dimensional minimal gauged supergravity \cite{Kerimo:2004gz}.
In (\ref{xi}), $\al$ is an arbitrary real integration constant, which turns out to be the only physical parameter, since by an appropriate diffeomorphism of the form (\ref{diff}), explicitely given in \cite{Cacciatori:2004rt}, it is possible to put the metric in a canonical form where $\F=\dd u\wedge\dd x$ and
\begin{equation}
  \G_{\al}(x,y,u)=-\frac{x^4}{\ell^2}+2\al
  x^3-{\al^2\ell^2}\left(x^2+y^2\right)\,,\qquad
  \varphi(u)=u\,. \label{Galpha}
\end{equation}
Finally, it is important to distinguish between the $\al=0$ and
$\al\neq0$ solutions. Siklos obtained a beautiful classification for spacetimes of the form (\ref{lobwave}), according to the number of independent Killing
vectors \cite{Siklos:1985}. It follows that for $\al=0$ the spacetime
admits a five-dimensional group of isometries, generated by five
Killing vectors. We shall meet again this metric as the timelike solution (\ref{el2}) of section \ref{electric}.
On the other hand, when $\al\neq0$, the canonical form of $\G$ falls
in the $A(x,y)$ class of \cite{Siklos:1985}, meaning that the only Killing vectors are $\p_u$ and $\p_v$.

In any case, by computing the norm squared of the Killing vectors $V$
constructed from the Killing spinors, one checks that
every one half supersymmetric lightlike solution is also a timelike solution.
\begin{figure}
	\begin{center}
    \begin{tabular}{|c||c|c|}
      \hline
      SUSY &
      \parbox[c]{5.5cm}{
        \centerline{purely gravitational
      solutions$\vphantom{\displaystyle1^{\displaystyle H}}$}
        \centerline{$\varphi'(u)=0\vphantom{\displaystyle\frac{.}{1}}$}
      } &
      \parbox[c]{5cm}{
        \centerline{electrovac spacetimes$\vphantom{\displaystyle1^{\displaystyle H}}$}
        \centerline{$\varphi'(u)\neq0\vphantom{\displaystyle\frac.1}$}
      } \\
      \hline
      \hline
      1/4     & Lobatchevski wave & $\G\neq\G_\al$       \\
      \hline
      1/2     & \textit{none}              & $\G_\al(u,x,y)$ \hspace{0.2cm} (\ref{Galpha}) \\
      \hline
      3/4     & \textit{none}              & \textit{none}        \\
      \hline
      Max  & AdS$_4$           & \textit{none}            \\
      \hline
    \end{tabular}
    \end{center}
  \caption{Classification of supersymmetric spacetimes in the lightlike
    case. Note that the fraction 3/4 of supersymmetry cannot be preserved.}
  \label{table_lightlike}
\end{figure}

The complete classification of supersymmetric solutions in the lightlike case is summarized in table \ref{table_lightlike}.
% we summarize the complete classification of supersymmetric solutions of the lightlike case.

%%%%%%%%%%%%%%%%%%%%%%%%%%%%%%%%%%%%%%%%%%%%%%%%%%%%%%%%%%%%%%%%%%%%%%%%%%%
\section{The timelike case}\label{timelikecase}

The general BPS solution for timelike $V^\mu$ reads \cite{Caldarelli:2003pb}
\begin{eqnarray}
\dd s^2 &=& -\frac 4{\ell^2 F\bar F}\left(\dd t + \omega_i \dd x^i\right)^2 + \frac{\ell^2 F\bar F}4
       \left[\dd z^2 + e^{2\phi}\left(\dd x^2 + \dd y^2\right)\right]\,, \label{metric} \\
{\cal F} &=& \frac{\ell^2}4 F\bar F\left[V\wedge\dd f+\dual\left(V\wedge\left(\dd
             g+\frac1\ell \dd z\right)\right)\right]\,, \nonumber
\end{eqnarray}
where $i=1,2$; $x^1 = x$, $x^2=y$, and we defined $\ell F=2i/(f-ig)$,
with $f=\bar{\epsilon}\epsilon$ and $g=i\bar{\epsilon}\Gamma_5\epsilon$.
The timelike Killing vector is given by $V = \partial_t$ and the functions $\phi$, $F$, $\bar F$, that depend on $(x,y,z)$, are determined by
the system
\begin{eqnarray}
\Delta F + e^{2\phi}[F^3 + 3FF' + F''] &=& 0\,, \label{F} \\
\Delta \phi + \frac 12 e^{2\phi}[F' + \bar{F}' + F^2 + \bar{F}^2 - F\bar F] &=& 0\,,
\label{phi} \\
\phi' - \re (F) &=& 0\,, \label{ReF}
\end{eqnarray}
where $\Delta = \partial^2_x + \partial^2_y$, and a prime denotes differentiation
with respect to $z$. (\ref{F}) comes from the combined Maxwell equation and Bianchi identity, whereas (\ref{phi}) results from the integrability condition for the Killing spinor $\epsilon$. Finally, the shift vector $\omega$ is obtained from
\begin{eqnarray}
\partial_z \omega_i &=& \frac{\ell^4}8 (F\bar F)^2 \epsilon_{ij}(f\partial_j g
- g\partial_j f)\,, \nonumber \\
\partial_i \omega_j - \partial_j \omega_i &=& \frac{\ell^4}8 (F\bar F)^2 e^{2\phi}
\epsilon_{ij}\left(f\partial_z g - g\partial_z f + \frac{2f}{\ell}\right)\,, \label{omega}
\end{eqnarray}
with $\epsilon_{12} = 1$.

%%%%%%%%%%%%%%%%%%%%%%%%%
Before presenting the explicit solutions of the timelike case, let us study some
general properties of the system (\ref{F}) -- (\ref{ReF}).

Decomposing $F$ into its real and imaginary part, $F = A + iB$, one checks that the real part of eqn.~(\ref{F}) follows from (\ref{phi}) and (\ref{ReF}),
so that the remaining system is
\begin{eqnarray}
\Delta B + e^{2\phi}\left[3\phi^{\prime 2} B - B^3 + 3\phi' B' + 3B\phi'' + B''\right] &=& 0\,, \label{B} \\
\Delta \phi + \frac 12 e^{2\phi}\left[2\phi'' + {\phi'}^2 - 3B^2\right] &=& 0\,, \label{phinew}
\end{eqnarray}
together with $A = \phi'$.
These equations can be derived from the action
\begin{eqnarray}
S = \int \dd^2x\,\dd z\,\left[\nabla B\cdot\nabla \phi + \frac 12 e^{2\phi}
\left(B^3 + 2B'\phi' + 3B{\phi'}^2\right)\right]\,,
    \label{actionBphi}
\end{eqnarray}
which enjoys various symmetries: first of all, we note that it is invariant under PSL$(2,\bR)$ transformations
\begin{equation}
z \to \frac{az + b}{cz + d}\,, \qquad ad - bc = 1\,, \label{PSLz}
\end{equation}
if $B$ transforms like a conformal field of weight two and $\phi$ as a Liouville field. As a consequence, $F$ has a connection-like transformation behaviour.
This symmetry acts in a nontrivial way on supergravity solutions, and can be used to generate new BPS solutions starting from known ones.

There is an additional infinite-dimensional conformal symmetry, corresponding to holomorphic coordinate transformations in $x+iy$. However, it is easy to see that, from the four-dimensional point of view, these represent diffeomorphisms that preserve the conformal gauge for the two-metric $e^{2\phi}(dx^2 + dy^2)$. Thus, unlike the PSL$(2,\bR)$ transformations above, this symmetry cannot be used to generate new solutions.

%%%%%%%%%%%%%%%%%%%%%%%%%%%%%%%%%%%%%%%%%%%%%%%%%%%%%%%%%%%%%%%%%%%%%%%%%%%%%
\subsection{``Purely magnetic'' solutions ($F$ real)}
\label{magnetic}

Let us consider first solutions with $F$ real. Then, in the coordinates (\ref{metric}), the electric components of the field $\mathcal F$ vanish. For this reason, we shall refer to these configurations as ``purely magnetic'' solutions.
One has $B=0$, and the only equation to solve is (\ref{phinew}), that reduces to
\begin{equation}
\Delta \phi + \frac 12 e^{2\phi}\left[{\phi'}^2 + 2\phi''\right] = 0\,.
\label{toda}\end{equation}
This is similar to the continuous $(SU(\infty))$ Toda equation (or heavenly equation) which determines self-dual Einstein metrics admitting at least one rotational Killing vector \cite{Boyer:mm}.

This highly nonlinear equation is difficult to solve; however, under the simplifying assumption that $A$ is a function of $z$ alone, one finds
\begin{equation}
A=\frac{2az+b}{az^2+bz+c}\,,\qquad
\phi = \int A(z)\,\dd z + \gamma(x,y)\,,
\label{Aphi}\end{equation}
with $a$, $b$ and $c$ being real integration constants obeying $4ac-b^2 = k$. Equation (\ref{toda}) reduces then to a Liouville equation for $\gamma(x,y)$,
\begin{equation}
  \Delta\gamma + \frac k2 e^{2\gamma} = 0\,, \label{Liouville}
\end{equation}
describing the metric on euclidean two-manifolds of constant curvature $k$.

When $a=0$, this curvature is negative, and one obtains the anti-Nariai AdS$_2\times\HH^2$ spacetime, with a purely magnetic Maxwell field,
\begin{eqnarray}
    \dd s^2 &=& -\frac{4z^2}{\ell^2}\dd t^2+\frac{\ell^2}{4z^2}
             \dd z^2 + \frac{\ell^2}{2x^2}(\dd x^2 + \dd y^2), \\
    {\cal F} &=& -\frac{\ell}{2x^2} \dd x \wedge \dd y\,.\nonumber
\end{eqnarray}
This configuration preserves half of the supersymmetries, and admits an
$osp(2|2)\oplus so(2,1)\cong su(1,1|1)\oplus su(1,1)$ isometry superalgebra \cite{Cacciatori:1999rp}.

When $a\neq0$, the supergravity solution is given by
\begin{eqnarray}
\dd s^2 &=& \left(\frac{z}{\ell} + \frac{k\ell}{4z}\right)^2 \dd t^2 +
         \frac{\dd z^2}{\left(\frac{z}{\ell} + \frac{k\ell}{4z}\right)^2} +
         z^2 e^{2\gamma}(\dd x^2 + \dd y^2)\,, \label{magnmonop} \\
{\cal F} &=& \frac{k\ell}4 e^{2\gamma} \dd x \wedge \dd y\,. \nonumber
\label{}\end{eqnarray}
and its properties are determined by the sign of the curvature $k$ of the transverse two-metric. We obtain thus the following classification:

\paragraph{Positive curvature:}
Using polar coordinates $(r, \sigma)$ on the $(x,y)$ plane, the metric and gauge field are given by (\ref{magnmonop}), with the Liouville field
\begin{equation}
e^{2\gamma} = \frac{2m^2}{kr^2 \cosh^2(m\ln r)}\,.
\end{equation}
Introducing new coordinates $(\theta,\varphi)$ defined by $r^m = \tan\theta/2$,
$m\sigma = \varphi$, we get
\begin{equation}
e^{2\gamma}(\dd r^2 + r^2 \dd\sigma^2) = \frac 2k (\dd\theta^2 + \sin^2\theta \dd\varphi^2)\,.
\end{equation}
As $\sigma$ is identified modulo $2\pi$, we see that $\varphi \sim \varphi + 2\pi m$, so that the effect of the parameter $m$ is to introduce a conical singularity ($0 < m < 1$) or an excess angle ($m > 1$) on the north and south poles of the two-sphere.
For $m=1$ there is no singularity, and the solution reduces to the one-quarter
supersymmetric magnetic monopole found by Romans \cite{Romans:1991nq}.
The magnetic charge of the solution reads
\begin{equation}
Q_m = \frac{1}{4\pi}\int {\cal F} = \frac{m\ell}2\,.
\end{equation}
If $m \neq 1$, there is a magnetic fluxline that passes through $x = y =0$,
which causes the magnetic charge to be different from the value $Q_m = \ell/2$
of Romans' solution.

\paragraph{Vanishing curvature:}
If $k=0$, the function $\gamma(x,y)$ is harmonic, so that the two-manifold
with metric $e^{2\gamma}(dx^2 + dy^2)$ is flat. The choice $\gamma = 0$ leads
to the maximally supersymmetric AdS$_4$ vacuum solution, which is 
the only configuration with maximal supersymmetry \cite{Caldarelli:2003pb}.

\paragraph{Negative curvature:}
For $a\neq0$ we can take without loss of generality
\begin{equation}
A = \frac{2z}{z^2 + \frac{k\ell^2}4}\,.
\end{equation}
Then, the solutions of the Liouville equation (\ref{Liouville}) are
classified according to their monodromy (see for example \cite{Seiberg:1990eb}).
Using polar coordinates $(r, \sigma)$ on the $(x,y)$ plane,
we have the following $\sigma$-independent solutions:\\
$
\begin{array}{@{}l@{\qquad}l}
\\
{\quad\ \,\bullet\ \,}\textrm{Elliptic monodromy:}
&\displaystyle e^{2\gamma} = -\frac{2m^2}{kr^2 \sinh^2(m\ln r)}\,,\\
\\
{\quad\ \,\bullet\ \,}\textrm{Parabolic monodromy:}
&\displaystyle e^{2\gamma} = -\frac{2}{kr^2 \ln^2 r}\,,\\
\\
{\quad\ \,\bullet\ \,}\textrm{Hyperbolic monodromy:}
&\displaystyle e^{2\gamma} = -\frac{2m^2}{kr^2 \sin^2(m\ln r)}\,.\\
\\
\end{array}
$\\
Here, $m$ is a constant related to the Liouville momentum.
In the case of elliptic monodromy and $m=1$, this reduces to the configuration
found in \cite{Caldarelli:1998hg}\footnote{For generalizations to five dimensions see \cite{Chamseddine:1999xk,Klemm:2000nj}.}, which preserves one quarter of the supersymmetry and admits an $s(2) \oplus su(1,1)$ isometry
superalgebra \cite{Cacciatori:1999rp}, and represents an extremal black hole with an hyperbolic event horizon located at $z = \sqrt{-k}\ell/2$.

%%%%%%%%%%%%%%%%%%%%%%%%%%%%%%%%%%%%%%%%%%%%%%%%%%%%%%%
\subsection{``Purely electric'' solutions ($F$ imaginary)}
\label{electric}

If one assumes instead the function $F$ to be purely imaginary, the field configurations have vanishing magnetic field in the coordinate system (\ref{metric}), and therefore we will refer to these solutions as the ``purely electric'' ones. In this case $A=0$, and thus $\phi$ and $B$ are independent of $z$. The equations (\ref{B}) and (\ref{phinew}) reduce to
\begin{eqnarray}
\Delta B - e^{2\phi}B^3 &=& 0\,, \label{Bg=0} \\
\Delta \phi - \frac 32 e^{2\phi}B^2 &=& 0\,. \label{phig=0}
\end{eqnarray}
This system follows from the two--dimensional dilaton gravity action
\begin{equation}
S = \int \dd^2 x \sqrt g \left[BR + B^3\right]\,, \label{dilgrav}
\end{equation}
if we use the conformal gauge $g_{ij}\dd x^i \dd x^j = e^{2\phi}(\dd x^2 + \dd y^2)$.
However, the equations of motion following from (\ref{dilgrav}) contain
also the constraints $\delta S/\delta g^{ij} = 0$ (whose trace yields
(\ref{Bg=0})), and therefore they are more restrictive than the system
(\ref{Bg=0}), (\ref{phig=0}). Of course, every solution that extremizes
the action (\ref{dilgrav}) is a solution of our system, but not vice versa.
It is interesting to note that (\ref{dilgrav}) is similar to the action
that arises from Kaluza-Klein reduction of the three-dimensional gravitational
Chern-Simons term \cite{Guralnik:2003we}, with the only difference that here
$B$ is a fundamental field, whereas the field arising in \cite{Guralnik:2003we}
is the curl of a vector potential.

%%%%%%%%%%%%%%%%%%%%%%%%%%
If we use the ansatz $B = x^{\kappa}$, equations
(\ref{Bg=0}) and (\ref{phig=0}) are satisfied for $\kappa = 2$ and
$\kappa = -1/3$. The former value of $\kappa$ yields the BPS
solution \cite{Caldarelli:2003pb} 
\begin{equation}
  \dd s^2=-\frac{\ell^4}{x^4}\left(\dd t+\frac{2x}\ell\,\dd y\right)^2
  +\frac{x^4}{\ell^4}\,\dd z^2+\frac{\ell^2}{2x^2}
  \left(\dd x^2+\dd y^2\right)\,,
\label{el1}\end{equation}
\begin{equation}
  {\mathcal F}=\frac{2\ell^2}{x^3}\,\dd t\wedge\dd x
  -\frac{7\ell}{2x^2}\,\dd x\wedge\dd y\,.
\end{equation}
This metric has four Killing vectors acting transitively on the whole spacetime and represents a homogeneous, stationary and geodesically complete BPS spacetime endowed with a nonnull electromagnetic field.
In the context of Einstein-Maxwell theory, this Petrov type I solution was obtained in \cite{Ozsvath}.

%%%%%%%%%%%%%%%%%%%%%%%%%%
For the other value of the exponent, $\kappa=-1/3$, the four-dimensional geometry and the electromagnetic field strength are respectively
\begin{equation}
\dd s^2 = -\frac{4x^{2/3}}{\ell^2}\left(\dd t - \frac{\ell^2}{6x^{4/3}}\dd y\right)^2
       + \frac{\ell^2}{4x^{2/3}}\dd z^2 + \frac{\ell^2}{9x^2}(\dd x^2 + \dd y^2)\,,
\label{el2}\end{equation}
\begin{equation}
{\cal F} = -\frac{2}{3\ell x^{2/3}}\dd t \wedge \dd x\,.
\end{equation}
By means of a coordinate transformation, this solution can be recast in the electrovac AdS travelling wave of section \ref{lightlike} with $\alpha=0$, and therefore preserves half of the supersymmetries and admits five Killing vectors.

%%%%%%%%%%%%%%%%%%%%%%%%%%%%%%%%%%%%%%%%%%%%%%%%%%%%%%%%%%%%%%%%%%%%%%%%%%%%%
\subsection{Solutions with $F=A(z)+iB(x,y,z)$.}

If we allow only for a $z$-dependence in $A$, the conformal
factor $\phi$ of the transverse two-metric is given by (\ref{Aphi}),
with $\gamma(x,y)$ an integration constant. Defining
\begin{equation}
\beta(x,y,z)=B\exp\int\!A(z)\,dz\,,
\label{defbeta}\end{equation}
the equations can be further simplified in two cases, according to which coordinates the function $\beta$ depends on.

%____________________________________________________________________________%
\subsubsection{The case $\beta=\beta(x,y)$}

The simplest case of vanishing $\beta$ has been analyzed in
section~\ref{magnetic}.
By relaxing this condition and allowing for an $(x,y)$-dependence in
$\beta$, the BPS configurations have $A$ and $\phi$ given by (\ref{phi}), and are described by the system
\begin{eqnarray}
\Delta \gamma + \frac 12 e^{2\gamma}\left(k - 3\beta^2\right) = 0\,,&
\label{gamma}\\
\vphantom{\frac12}\Delta \beta + e^{2\gamma}\left(k\beta - \beta^3\right) = 0\,,& \label{beta}
\end{eqnarray}
where $k$ denotes an arbitrary constant and $(a,b,c)$ are real integration constants obeying $4ac - b^2 = k$.

It is interesting to observe that if we choose $a=1$, $b = c = 0$, so that $k = 0$ and $A = 2/z$, the system for $\beta$ and $\gamma$ reduces to the same set of equations (\ref{phig=0}) and (\ref{Bg=0}) that describes ``purely electric'' solutions. Therefore, to any ``purely electric'' solution we can associate a new BPS configuration. Let us apply this to the solutions presented in section~\ref{electric}. Starting from the Petrov type I solution (\ref{el1}) one obtains
\begin{eqnarray}
\dd s^2 &=& -\frac{z^4}{\ell^2(z^2+x^4)}\dd t^2+\frac{\ell^2(z^2+x^4)}{z^4}\dd z^2
         + \frac{\ell^2(z^2+x^4)}{2x^6} \dd x^2 + \nonumber \\
&& + \frac{2z^2(z^2-6x^4)}{3x^3(z^2+x^4)} \dd t\,\dd y 
   + \frac{7\ell^2(z^4+6z^2x^4-9x^8)}{18x^6(z^2+x^4)} \dd y^2\,, \\
{\cal F} &=& \frac{1}{(z^2+x^4)^2}\left[-\frac{2}{\ell}xz^2
             (z^2-x^4)\dd t\wedge\dd x-\frac{2}{\ell}x^6z\,\dd t\wedge\dd z\;
             +\right.\nonumber \\
&& \left. + \frac{14}{3}\ell x^3 z\,\dd y \wedge \dd z
         - \frac{7\ell}{6x^2}(z^2-x^4)(z^2-3x^4)
   \dd x \wedge \dd y \right]\,, \nonumber
\end{eqnarray}
while the configuration related to the BPS solution (\ref{el2}) is
\begin{eqnarray}
\dd s^2 &=& -\frac{z^4x^{2/3}}{\ell^2(1+z^2x^{2/3})} \dd t^2 + \frac{2z^2}{3x^{2/3}} \dd t\,
         \dd y + \frac{\ell^2(1+z^2x^{2/3})}{z^4x^{2/3}} \dd z^2 \nonumber \\
&& + \frac{\ell^2(1+z^2x^{2/3})}{9x^2} \dd x^2\,, \\
{\cal F} &=& -\frac{2zx^{1/3}}{\ell(1+z^2x^{2/3})^2} \dd t \wedge \dd z -
             \frac{z^2(1-z^2x^{2/3})}{3\ell x^{2/3}(1+z^2x^{2/3})^2} \dd t\wedge \dd x\,. \nonumber
\end{eqnarray}
A calculation of the Weyl scalars shows that the two spacetimes, as before,
are of Petrov type I and N respectively. Note that these solutions can also be obtained from their ``purely electric'' counterparts by an appropriate PSL$(2,\bR)$ transformation.

%____________________________________________________________________________%
\subsubsection{Kink solutions and generalizations}
%%%%%%%%%%%%%%%%%%%%%%%%%%%%%%%%%%%%%

More general solutions can be obtained in the $\beta=\beta(x,y)$ case.
As shown in \cite{Cacciatori:2004rt}, the system (\ref{gamma})-(\ref{beta})
follows from the dimensionally reduced gravitational Chern-Simons
model considered in \cite{Guralnik:2003we}. Therefore, to any solution of this model we can associate a BPS solution of the gauged supergravity. Using (for $k>0$) the ``kink'' solution \cite{Guralnik:2003we}, one obtains the supergravity solution
\begin{eqnarray}
\dd s^2 &=& -\frac 1{\ell^2}\frac{(z^2 + \frac{\ell^2}4)^2}{z^2 + \frac{\ell^2}4
         \tanh^2\frac X2}\left[\dd t + \left(\frac{\ell^3}{4(z^2 + \frac{\ell^2}4)
         \cosh^4\frac X2} - \frac{\ell}{\cosh^2\frac X2}\right)\dd y\right]^2 \nonumber \\
     & & + \ell^2\frac{z^2 + \frac{\ell^2}4\tanh^2\frac X2}{(z^2 + \frac{\ell^2}4)^2}\dd z^2
         + (z^2 + \frac{\ell^2}4\tanh^2\frac X2)(\dd X^2 + \frac{\dd y^2}{\cosh^4\frac X2})\,,
         \label{kink} \\
{\cal A} &=& \frac 12 \frac{z^2 + \frac{\ell^2}4}{z^2 + \frac{\ell^2}4\tanh^2\frac X2}
             \tanh\frac X2\,\dd t\,. \nonumber
\end{eqnarray}
Asymptotically for $X \to \pm \infty$ the gauge field goes to zero and the
metric approaches AdS$_4$ written in nonstandard
coordinates \cite{Alonso-Alberca:2000cs}, so that the ``kink''
solution (\ref{kink}) interpolates between two AdS vacua at
$X=\pm\infty$.

Grumiller and Kummer were able to write down the most general solution
of the dimensionally reduced gravitational Chern-Simons theory using the fact that it can be written as a Poisson-sigma model with four-dimensional target space and degenerate Poisson tensor of rank two \cite{Grumiller:2003ad}.
This gives rise to new BPS supergravity solutions generalizing (\ref{kink}),
\begin{eqnarray}
\dd s^2 &=& -\frac 1{\ell^2}\frac{(z^2 + \frac{\ell^2}4)^2}{z^2 + \frac{\ell^2}4
         \tanh^2\frac X2}\left[\dd t + \left(\frac{\ell^3(1+\delta)}{4(z^2 + \frac{\ell^2}4)
         \cosh^4\frac X2} - \frac{\ell}{\cosh^2\frac X2}\right)\dd y\right]^2 \nonumber \\
     & & + \ell^2\frac{z^2 + \frac{\ell^2}4\tanh^2\frac X2}{(z^2 + \frac{\ell^2}4)^2} \dd z^2
         +\frac {(z^2 + \frac{\ell^2}4\tanh^2\frac X2)}{1+\delta}\left(\dd X^2 + \frac{(1+\delta)^2}
         {\cosh^4\frac X2} \dd y^2 \right)\,,
         \nonumber \\
{\cal A} &=& \frac 12 \frac{z^2 + \frac{\ell^2}4}{z^2 + \frac{\ell^2}4\tanh^2\frac X2}
             \tanh\frac X2\,\dd t +\delta_0 \frac {\ell^3}8
             \frac {\tanh\frac X2}{z^2+\frac {\ell^2}4 \tanh^2\frac X2} \,\dd y \,. \label{kink1}
\end{eqnarray}
with $\delta=\delta_0\cosh^4\frac X2$.
In the special case $\delta_0=0$ we recover the kink solution considered above. This solutions has been recently further analyzed in \cite{Bergamin:2004me}, where it was interpreted as a soliton consisting of photons kept together by gravity.

%If $\delta_0 \leq 0$ then the metric is well-defined only in the region
%$-\bar X \leq X \leq \bar X$ where $\cosh({\bar X}/2) = -1/\delta_0$
%and $\delta_0 \geq -1$.\\
%If $\delta_0 \geq 0$ we can take $X \to \pm \infty$ so that the metric becomes
%\begin{displaymath}
%\dd s^2 = -\left(\frac{z^2}{\ell^2} + \frac 14\right)\left[\dd t \mp \frac{\ell}u \dd y
%+\frac {l^3 \delta_0}{4z^2 +\ell^2} \dd y \right]^2
%         + \frac{\dd z^2}{\frac{z^2}{\ell^2} + \frac 14} +
%         \ell^2\left(\frac{z^2}{\ell^2}
%         + \frac 14\right)\left[ \frac {\dd u^2}{\delta_0 u^4} +\delta_0\dd y^2 \right]\,,
%\end{displaymath}
%where again $u = \pm e^{\pm X}/4$,
%and the gauge field asymptotes to
%\begin{equation}
%{\cal A} =\frac {\delta_0 \ell}8 \frac {\dd y}{\frac{z^2}{\ell^2} + \frac 14}.
%\end{equation}

Note that all these solutions are defined only for $k\geq 0$. One
can now verify that the domain of the parameter $k$ can be
extended also to the negative region. This yields the solution
\begin{eqnarray}
\dd s^2 &=& -\frac 1{\ell^2}\frac{(z^2 - \frac{\ell^2}4)^2}{z^2 +
\frac{\ell^2}4
         \tan^2\frac X2}\left[\dd t - \left(\frac{\ell^3(1+\delta)}{4(z^2 - \frac{\ell^2}4)
         \cos^4\frac X2} + \frac{\ell}{\cos^2\frac X2}\right)\dd y\right]^2 \nonumber \\
     & & + \ell^2\frac{z^2 + \frac{\ell^2}4\tan^2\frac X2}{(z^2 - \frac{\ell^2}4)^2} \dd z^2
         +\frac {(z^2 + \frac{\ell^2}4\tan^2\frac X2)}{1+\delta}\left(\dd X^2 + \frac{(1+\delta)^2}
         {\cos^4\frac X2} \dd y^2 \right)\,,
         \nonumber \\
{\cal A} &=& -\frac 12 \frac{z^2 - \frac{\ell^2}4}{z^2 +
\frac{\ell^2}4\tan^2\frac X2}
             \tan\frac X2\,\dd t +\delta_0 \frac {\ell^3}8
             \frac {\tan\frac X2}{z^2+\frac {\ell^2}4 \tan^2\frac X2} \,\dd y \,, \label{kink1-neg} 
\end{eqnarray}
with $\delta = \delta_0\cos^4\frac{X}2$.

%____________________________________________________________________________%
\subsubsection{The case $\beta=\beta(z)$}
\label{betazeta}

In this case $B$ must be a function of the coordinate $z$
alone, and the system of equations describing this subset of
solutions is
\begin{eqnarray}
B''+3\left(AB\right)'+B\left(3A^2-B^2\right)&=&0\,,\\
\check\RR(\gamma)&=&k\,,\label{ll}\\
e^{2\int\dd zA(z)}\left[2A'+A^2-3B^2\right]&=&k\,,
\end{eqnarray}
where $\check\RR(\gamma)$ is the curvature scalar of the two-manifold with metric $\dd
s^2=e^{2\gamma}\left(\dd x^2+\dd y^2\right)$. Without loss of generality we can set $k=0,\pm1$. The general solution for the complex function $F=F(z)$ is therefore \cite{Caldarelli:2003pb}
\begin{equation}
    F=\frac{2az+b}{az^2+bz+c}\,,
\end{equation}
where $a$, $b$ and $c$ are complex integration constants (if these
constants are real, we fall back in the ``purely magnetic'' case
already considered in section~\ref{magnetic}).
If we take $a=0$, we obtain the solution AdS$_2\times\HH^2$ with magnetic flux on $\HH^2$ already considered in section \ref{magnetic}.

The case $a\neq0$ was solved in \cite{Caldarelli:2003pb}, where it was shown that one recovers the supersymmetric Reissner-Nordstr\"om-Taub-NUT-AdS$_4$ solutions \cite{Alonso-Alberca:2000cs} with metric
\begin{eqnarray}
\displaystyle  ds^2 = N(dt + \omega_{\phi} d\phi)^2 - \frac{dz^2}{N}
  + (z^2 + n^2) (d\theta^2 + S_k^2(\theta) d\phi^2)\,,&
       \label{RNTNAdS}\\
  N = -\frac{\vphantom{\frac12}z^4 + 2z^2(n^2 - lP) - 4nlQz + n^2(n^2 + 2lP)
  + l^2(Q^2 + P^2)}{l^2(z^2 + n^2)\vphantom{\frac12}}\,,&
        \label{lapse}\\
 \omega_{\phi} = \left\{\begin{array}{l}
                     -2n \theta \\ 2n \cos\theta \\
                     -2n \cosh\theta
                     \end{array} \right.
\quad\mathrm{and\ \ }
S_k(\theta) = \left\{\begin{array}{l}
                     1\\ \sin\theta\\ \sinh\theta
                     \end{array} \right.
\quad\mathrm{for\ \ }
k=\left\{\begin{array}{l} 0 \\ 1\\ -1 \end{array}\right..&
\end{eqnarray}
Here, $n$ denotes the NUT parameter, and $Q$ and $P$ are the electric and
magnetic charges respectively. These are fixed by the constants $a$, $b$ and $c$,
\begin{eqnarray*}
n &=& i\left(\frac{\bar b}{4\bar a} - \frac{b}{4a}\right)\,, \\
Q &=& \frac{i}{2l}\left(\frac ca - \frac{\bar c}{\bar a}\right)\,, \\
P &=& -\frac{1}{2l}\left(2n^2 + \frac ca + \frac{\bar c}{\bar a}\right)\,.
\end{eqnarray*}
The electromagnetic field strength is easily obtained from (\ref{F}). One checks that the final solution belongs to the Reissner-Nordstr\"om-Taub-Nut-AdS class of solutions \cite{Alonso-Alberca:2000cs}, with arbitrary nut charge $n$ and electric charge $Q$, whereas the magnetic charge $P$ and mass parameter $M$ are given by
\begin{equation}
  P = -\frac{kl^2 + 4n^2}{2l}\,, \qquad M = \frac{2nQ}{l}\,. \label{susycond}
\end{equation}
These are exactly the conditions on $P$ and $M$ found in
\cite{Alonso-Alberca:2000cs}, under which the RN-TN-AdS solutions preserve one
quarter of the supersymmetry\footnote{The other sign for $P$ given in
  \cite{Alonso-Alberca:2000cs} can be obtained by replacing $\phi \to -\phi$,
$n \to -n$.}.
Note that, taking more general solutions of the Liouville equation (\ref{ll}), one would obtain a larger class of BPS configurations, and in particular recover all purely magnetic solutions of section \ref{magnetic}.

%%%%%%%%%%%%%%%% HARMONIC SOLUTIONS %%%%%%%%%%%%%%%%%%%%%%%%%%%%%%%%%%%%
\subsection{Harmonic solutions}
Another rich class of solutions can be found if we choose $F$ to be
harmonic, $\Delta F = 0$. Equation (\ref{F}) gives
\begin{equation}
F =\frac {2a z +b}{az^2 +bz +c}\,,
\label{Fharm}\end{equation}
where $a$, $b$ and $c$ are arbitrary holomorphic functions of $\zeta=x+iy$.
Using the Eddington-Finkelstein-like coordinate
$u=t+\frac {\ell^2}4 \int F\bar F \dd z$, the solution assumes the form
\begin{equation}
  \dd s^2 = -\frac4{\ell^2F\bar F}\left(\dd u+w\right)^2
  +2\left(\dd u+w\right)\dd z
  +\frac {\ell^2}4 |2az+b|^2 e^{2\gamma} \left(\dd x^2 +\dd y^2
  \right),
\end{equation}
\begin{equation}
 {\mathcal F}=\dd{\mathcal A}\,,\qquad
 {\mathcal A} =\frac i{\ell}\left( \frac 1F -\frac 1{\bar F} \right) (\dd u+w) +\frac {\ell}2 \check\dd\gamma \,. 
\end{equation}
Here, $\check\dd\equiv\dd x^i\epsilon_{ij}\partial_j$, and $w=\check\dd\psi$, where
$\psi (x,y)$ is a function satisfying
\begin{eqnarray}
\Delta \psi = i \frac {\ell^2}2 e^{2\gamma} (a\bar b -b\bar a)\,.
\label{psi}\end{eqnarray}
Finally
\begin{equation}
\check\RR\left(\gamma\right)=2\left(a\bar c+\bar ac\right) -b\bar b\,,
\label{gammaharm}\end{equation}
showing that the scalar curvature
$\check\RR\left(\gamma\right)=-e^{-2\gamma}\Delta2\gamma$ of the
two-dimensional metric $e^{2\gamma} (\dd x^2 +\dd y^2 )$ is not
constant in general.

Let us consider now some particular solutions. If the functions $a$,
$b$ and $c$ are constant, we fall in the cases already studied in
sections~\ref{magnetic} and \ref{betazeta}. More precisely, if $a=0$
we obtain the anti-Nariai spacetime AdS$_2\times\HH^2$, while for
$a\neq0$ the BPS limits of the RNTN-AdS$_4$ family of solutions are
recovered. We will analyse now these two cases allowing for
non-constant functions.

%___________________________________________________________________________%
\subsubsection{Supersymmetric Kundt solutions ($a=0$)}

When $a=0$, after performing some diffeomorphisms, one finds that the transverse two-metric has a constant negative curvature, and describes (at least locally) an
hyperbolic plane $\HH^2$. The metric and gauge field read
\begin{eqnarray}
    \dd s^2&=&-\frac{4}{\ell^2}|z+c|^2\dd u^2 +2\dd u\dd
    z+\frac{\ell^2}{2x^2}\left(\dd x^2+\dd y^2\right)\,,\\
    {\cal A}&=&\frac{i}{\ell}\left(c-\bar c\right)\dd
    u+\frac{\ell}{2x}\dd y\,,
\end{eqnarray}
where $c$ is an arbitrary holomorphic function $c=c(\zeta)$.
This metric is precisely of the Kundt form, describing manifolds
admitting a non-expanding, non-twisting null congruence of geodesics,
and the solution has the interpretation of {\em supersymmetric
  electromagnetic and gravitational waves propagating on anti-Nariai
  spacetime}. This is a particular case of the more general solution
found in \cite{Podolsky:2002sy}. Its Petrov type is II.

%___________________________________________________________________________%
\subsubsection{Supersymmetric Robinson-Trautman solutions ($a\neq0$)}

If, instead, $a\neq0$, the solution is determined by two arbitrary functions $\beta(\zeta)$ and $\delta(\zeta)$, holomorphic in $\zeta=x+iy$. The solution reads
\begin{equation}
    \dd s^2=-\frac{\left|(z+\beta)^{2}-\delta\right|^2}{\ell^2|z+\beta|^{2}}\left(\dd u+w\right)^{2} 
    +2\left(\dd u+w\right)\dd  z+\ell^{2}|z+\beta|^{2}e^{2\gamma}\left(\dd x^2+\dd y^2\right)\,,\\
\end{equation}
\begin{equation}
    {\cal A}=-\frac{\im\left[\left((z+\beta)^{2}-\delta\right)(z+\bar{\beta})\right]}
    {\ell\left|z+\beta\right|^{2}}\left(\dd u+w\right)+\frac{\ell}{2}\check\dd\gamma\,.
\label{rtgen}\end{equation}
The system of equations for $\gamma$ and $\psi$ describing this class of supersymmetric configurations is
\begin{eqnarray}
    \check\RR(\gamma) &=& -4\left[2(\im\,\beta)^2+\re\,\delta\right]\,,\\
    \Delta\psi &=& 2\ell e^{2\gamma}\im\,\beta\,.\label{rtpsi}
\end{eqnarray}
If $\beta$ and $\delta$ are constant functions, then, as already
stressed, this solution is of Petrov type D and belongs to the
RNTN-AdS$_4$ family of solutions.
In general, solution (\ref{rtgen}) describes
electromagnetic and gravitational expanding waves propagating on these
supersymmetric RNTN-AdS$_4$ backgrounds.

The simplest such solution is obtained taking $\beta$ to be some real constant $\kappa$. Then $w=0$, and defining the function $P=\sqrt{2}\ell^{-1}e^{-\gamma}$ and the operator
$\Delta^{*}\equiv\frac 12P^{2}\Delta$, the solution reads
\begin{eqnarray}
    \dd s^2=-\left[\frac{z^{2}}{\ell^{2}}+\Delta^{*}\ln P+\frac{|\delta|^{2}}
    {\ell^2z^{2}}\right]\dd u^{2} 
    +2\dd u\dd  z+\frac{2z^{2}}{P^{2}}\left(\dd x^2+\dd y^2\right)\,,\\
    {\cal A}=\frac{\im\,\delta}{\ell z}\,
    \dd u-\frac{\ell}{2}\check\dd\ln P\,.
    \qquad\qquad\qquad\qquad\qquad\qquad\qquad\qquad\quad\,
\end{eqnarray}
with $P(x,y)$ any solution of $\Delta^{*}\Delta^{*}\ln P=0$, while
$\im\,\delta$ is determined by the fact that $\delta$ is holomorphic
and its real part is fixed by $\check\RR(\gamma)=-4\,\re\,\delta$.
This metric is clearly of the Robinson-Trautman form, describing
manifolds admitting an expanding, non-twisting null congruence of
geodesics. In fact, it is of Petrov type II and generalizes the massless and purely gravitational Robinson-Trautman class of solutions found in \cite{Podolsky:2002sy} by adding electromagnetic waves in it. It has therefore the interpretation of {\em supersymmetric electromagnetic and gravitational expanding waves propagating on the BPS Reissner-Nordstr\"om-AdS$_4$ spacetimes}.

%%%%%%%%%%%%%%%%%%%%%%%%%%%%%%%%%%%%%%%%%%%%%%%%%%%%%%%%%%%%%%%%%%%%%%%%%%%%
\section{Inclusion of external matter}
\label{external}

In \cite{Caldarelli:2003wh} the analysis of BPS solutions to minimal gauged supergravity in four dimensions was extended to include external matter.
As in the ungauged case \cite{Tod:pm}, supersymmetry imposes the condition that the sources form a perfect fluid with vanishing pressure. Such sources, together with a negative cosmological constant are the needed ingredients to obtain the G\"odel universe \cite{Godel:ga}. It turns out that there are members of the family of G\"odel-type solutions in four dimensions \cite{Reboucas:hn} that preserve one quarter of the supersymmetry \cite{Caldarelli:2003wh}.

If we allow external charged sources carrying electric current $J^e_\mu$ and
magnetic current $J^m_\mu$, the Maxwell equations read
\begin{equation}
\nabla^\mu{\cal F}_{\mu\nu}=-4\pi J^e_\nu\,,\qquad
\nabla_{[\mu}{\cal F}_{\nu\rho]}=\frac{4\pi}3\ep_{\mu\nu\rho}{}^\si J^m_\si\,.
\end{equation}
%The charged sources carry some energy-momentum $T^{\rm ext}_{\mu\nu}$. 
Imposing the integrability conditions (\ref{intcond}) one obtains, after some algebra, the BPS conditions on the currents,
\begin{equation}
J=J^e+iJ^m=\kappa(f-ig)V\,, \label{J}
\end{equation}
where $\kappa$ is an arbitrary real function. Conservation of $J$ requires time-independence of $\kappa$, and the Einstein equations imply that these sources have the dust stress tensor $T^{\rm ext}_{\mu\nu}=\kappa V_\mu V_\nu$.
%\begin{equation}
%T^{\rm ext}_{\mu\nu}=\kappa V_\mu V_\nu\,.
%\end{equation}
Conservation of the combined energy-momentum tensor of the electromagnetic field and the sources finally imposes the vanishing of magnetic current $J^m$, and thus $g=0$.

The general timelike solution is then found following the same steps as those leading to the sourceless timelike solution
%One can now follow closely the discussion of the sourceless timelike
%solutions 
\cite{Caldarelli:2003pb}. In fact, the only difference is in
Maxwell's equations which include now the sources. The general supersymmetric
solution with timelike Killing vector $V^\mu$ is therefore still given by
(\ref{metric}), (\ref{phi}) and (\ref{omega}), but with (\ref{F}) replaced by
\begin{equation}
\Delta F + e^{2\phi}[F^3 + 3FF' + F'' + 4\pi\kappa F] = 0\,.
\end{equation}
Furthermore, the condition $g=0$ yields $\bar F = -F$, $F' = 0$, so that now
the functions $F$ and $\phi$ are determined by solving the system
\begin{eqnarray}
\Delta F + e^{2\phi}[F^3 + 4\pi\kappa F] &=& 0\,, \label{Fmod} \\
\Delta \phi + \frac 32 e^{2\phi}F^2 &=& 0\,. \label{phimod}
\end{eqnarray}

A simple solution of these equations can been obtained in assuming that $\kappa$ is a positive constant. This leads to the metric
\begin{equation}
ds^2=-\frac1{\pi\kappa\ell^2}
          \left(dt-\frac{2\ell^2\sqrt{\pi\kappa}}{3x}\ dy\right)^2
          +\pi\kappa\ell^2\ dz^2
          + \frac{\ell^2}{6x^2}\left(dx^2 + dy^2\right)\,, \label{metric2}
\end{equation}
with a constant magnetic flux through the transverse manifold,
\begin{equation}
{\cal F} = \frac{\ell}{6x^2}\ dx\wedge dy\,.
\end{equation}
A detailed analysis shows that this solution preserves exactly $1/4$ of the
original supersymmetry and represents a G\"odel-type spacetime with vorticity $\Omega=2/\ell$ and parameter $m=\sqrt6/\ell$ \cite{Reboucas:hn}. This is a  spacetime homogeneous universe with rigidly rotating dust and a magnetic field.
%Rebou\c{c}as-Tiomno metric \cite{Reboucas:hn} 
%The presence of CTCs is most obvious in cylindrical coordinates,
%\begin{equation}
% ds^2=-\left(dt-\frac{4\Omega}{m^2}\sinh^2\left(\frac{mr}2\right)
% d\varphi\right)^2+\frac1{m^2}\sinh^2(mr)\ d\varphi^2+dr^2+dz^2\,.
%\end{equation}
%For sufficiently large $r$ the vector $\partial_\varphi$ is timelike, and
%its integral curves become CTCs.

%%%%%%%%%%%%%%%%%%%%%%%%%%%%%%%%%%%%%%%%%%%%%%%%%%%%%%%%%%%%%%%%%%%%%%%%%%

\section{Final remarks}
\label{final}

In conclusion, significant progress has been made in the understanding of supersymmetric solutions to minimal gauged supergravity in four dimensions. In the lightlike case, a complete description of the BPS configurations has been obtained, showing in particular an enhancement of the supersymmetry for very particular profiles of the traveling waves. The timelike case is much more complicated, but we have been able to get some insight in the mathematical structure underlying the equations governing it, and to extract different subfamilies of BPS solutions of physical interest, some of which were previously unknown.

However, in the timelike case, many issues remain to be settled. One would like to refine the classification in the sense of finding the additional restrictions on the geometries in order that they preserve more than one supersymmetry, and to see whether 3/4 supersymmetric solutions are possible at all. This is technically not easy, since the timelike solution is much less explicit than the lightlike one, but may uncover new interesting supergravity solutions, as for example configurations representing microstates for supersymmetric AdS black holes.
Moreover, it should be possible to recover the supersymmetric Kerr-Newman-AdS solution \cite{Kostelecky:1995ei,Caldarelli:1998hg} from the general form of the BPS solutions, but this probably involves some complicated coordinate change, and we did not yet succeed to obtain it.

A deeper understanding of the mathematical structure behind the system of differential equations governing the timelike solutions and of its symmetries would simplify these tasks.
Since a subsector of the general equations yields the dimensionally reduced gravitational Chern-Simons action, one may wonder whether the complete
system (\ref{F}) -- (\ref{ReF}) is described by the gravitational Chern-Simons
theory in three dimensions \cite{Deser:1981wh}.

Finally, one could consider matter-coupled gauged supergravity, where a large
class of supersymmetric black holes is known \cite{Sabra:1999ux,Chamseddine:2000bk}, and see whether a classification of BPS solutions is still feasible.

%%%%%%%%%%%%%%%%%%%%%%%%%%%%%%%%%%%%%%%%%%%%%%%%%%%%%%%%%%%%%%%%%%%%%%%%%%%%%%

\section*{Acknowledgments}

This work was partially supported by INFN, MURST and
by the European Commission RTN program
HPRN-CT-2000-00131, in which M.~M.~C.~is associated to the University of Torino.
\normalsize

\newpage

%%%%%%%%%%%%%%%%%%%%%%%%%%%%%%%%%%%%%%%%%%%%%%%%%%%%%%%%%%%%%%%%%%%%%%%%%%%%%%


\begin{thebibliography}{99}

%\cite{Tod:pm}
\bibitem{Tod:pm}
K.~P.~Tod,
``All Metrics Admitting Supercovariantly Constant Spinors,''
Phys.\ Lett.\ B {\bf 121}, 241 (1983).
%%CITATION = PHLTA,B121,241;%%

%\cite{Gauntlett:2002sc}
\bibitem{Gauntlett:2002sc}
J.~P.~Gauntlett, D.~Martelli, S.~Pakis and D.~Waldram,
``G-structures and wrapped NS5-branes,''
Commun.\ Math.\ Phys.\  {\bf 247}, 421 (2004)
[arXiv:hep-th/0205050].
%%CITATION = HEP-TH 0205050;%%

%\cite{sugra}
\bibitem{sugra}
J.~P.~Gauntlett, J.~B.~Gutowski, C.~M.~Hull, S.~Pakis and H.~S.~Reall,
``All supersymmetric solutions of minimal supergravity in five dimensions,''
Class.\ Quant.\ Grav.\  {\bf 20} (2003) 4587
[arXiv:hep-th/0209114];\\
%%CITATION = HEP-TH 0209114;%%
%\cite{Gauntlett:2002fz}
J.~P.~Gauntlett and S.~Pakis,
``The geometry of D = 11 Killing spinors,''
JHEP {\bf 0304} (2003) 039
[arXiv:hep-th/0212008];\\
%%CITATION = HEP-TH 0212008;%%
%\cite{Gauntlett:2003wb}
J.~P.~Gauntlett, J.~B.~Gutowski and S.~Pakis,
``The geometry of D = 11 null Killing spinors,''
JHEP {\bf 0312} (2003) 049
[arXiv:hep-th/0311112];\\
%%CITATION = HEP-TH 0311112;%%
%\cite{Gauntlett:2003cy}
J.~P.~Gauntlett, D.~Martelli and D.~Waldram,
``Superstrings with intrinsic torsion,''
Phys.\ Rev.\ D {\bf 69} (2004) 086002
[arXiv:hep-th/0302158];\\
%%CITATION = HEP-TH 0302158;%%
%\cite{Gutowski:2003rg}
J.~B.~Gutowski, D.~Martelli and H.~S.~Reall,
``All supersymmetric solutions of minimal supergravity in six dimensions,''
Class.\ Quant.\ Grav.\  {\bf 20} (2003) 5049
[arXiv:hep-th/0306235];\\
%%CITATION = HEP-TH 0306235;%%
%\cite{Cariglia:2004kk}
M.~Cariglia and O.~A.~P.~Mac Conamhna,
``The general form of supersymmetric solutions of N = (1,0) U(1) and SU(2)
gauged supergravities in six dimensions,''
arXiv:hep-th/0402055;\\
%%CITATION = HEP-TH 0402055;%%
%\cite{Cariglia:2004qi}
%\bibitem{Cariglia:2004qi}
M.~Cariglia and O.~A.~P.~Mac Conamhna,
``Timelike Killing spinors in seven dimensions,''
arXiv:hep-th/0407127.
%%CITATION = HEP-TH 0407127;%%

%\cite{Gauntlett:2003fk}
\bibitem{Gauntlett:2003fk}
J.~P.~Gauntlett and J.~B.~Gutowski,
``All supersymmetric solutions of minimal gauged supergravity in five
dimensions,''
Phys.\ Rev.\ D {\bf 68} (2003) 105009
[arXiv:hep-th/0304064].
%%CITATION = HEP-TH 0304064;%%

%\cite{Caldarelli:2003pb}
\bibitem{Caldarelli:2003pb}
M.~M.~Caldarelli and D.~Klemm,
``All supersymmetric solutions of $N = 2$, $D = 4$ gauged supergravity,''
JHEP {\bf 0309} (2003) 019
[arXiv:hep-th/0307022].
%%CITATION = HEP-TH 0307022;%%

%\cite{Gutowski}
\bibitem{Gutowski}
J.~B.~Gutowski and H.~S.~Reall,
``Supersymmetric AdS(5) black holes,''
JHEP {\bf 0402} (2004) 006
[arXiv:hep-th/0401042];\\
%%CITATION = HEP-TH 0401042;%%
J.~B.~Gutowski and H.~S.~Reall,
``General supersymmetric AdS$_5$ black holes,''
JHEP {\bf 0404} (2004) 048
[arXiv:hep-th/0401129].
%%CITATION = HEP-TH 0401129;%%

\bibitem{d1d5}
%\cite{Lunin:2004uu}
%\bibitem{Lunin:2004uu}
O.~Lunin,
``Adding momentum to D1-D5 system,''
JHEP {\bf 0404} (2004) 054
[arXiv:hep-th/0404006];\\
%%CITATION = HEP-TH 0404006;%%
%\cite{Giusto:2004id}
%\bibitem{Giusto:2004id}
S.~Giusto, S.~D.~Mathur and A.~Saxena,
``Dual geometries for a set of 3-charge microstates,''
Nucl.\ Phys.\ B {\bf 701} (2004) 357
[arXiv:hep-th/0405017];\\
%%CITATION = HEP-TH 0405017;%%
%\cite{Giusto:2004ip}
%\bibitem{Giusto:2004ip}
S.~Giusto, S.~D.~Mathur and A.~Saxena,
``3-charge geometries and their CFT duals,''
arXiv:hep-th/0406103.
%%CITATION = HEP-TH 0406103;%%

%\cite{Mathur:2003hj}
\bibitem{Mathur:2003hj}
S.~D.~Mathur, A.~Saxena and Y.~K.~Srivastava,
``Constructing 'hair' for the three charge hole,''
Nucl.\ Phys.\ B {\bf 680} (2004) 415
[arXiv:hep-th/0311092].
%%CITATION = HEP-TH 0311092;%%

%\cite{Lin:2004nb}
\bibitem{Lin:2004nb}
H.~Lin, O.~Lunin and J.~Maldacena,
``Bubbling AdS space and 1/2 BPS geometries,''
arXiv:hep-th/0409174.
%%CITATION = HEP-TH 0409174;%%

%\cite{Caldarelli:2003wh}
\bibitem{Caldarelli:2003wh}
M.~M.~Caldarelli and D.~Klemm,
``Supersymmetric G\"odel-type universe in four dimensions,''
Class.\ Quant.\ Grav.\  {\bf 21} (2004) L17
[arXiv:hep-th/0310081].
%%CITATION = HEP-TH 0310081;%%

%\cite{Cacciatori:2004rt}
\bibitem{Cacciatori:2004rt}
S.~L.~Cacciatori, M.~M.~Caldarelli, D.~Klemm and D.~S.~Mansi,
``More on BPS solutions of N = 2, d = 4 gauged supergravity,''
JHEP {\bf 0407} (2004) 061
[arXiv:hep-th/0406238].
%%CITATION = HEP-TH 0406238;%%

%\cite{Romans:1991nq}
\bibitem{Romans:1991nq}
L.~J.~Romans,
``Supersymmetric, cold and lukewarm black holes in cosmological
Einstein-Maxwell theory,''
Nucl.\ Phys.\ B {\bf 383} (1992) 395
[arXiv:hep-th/9203018].
%%CITATION = HEP-TH 9203018;%%

%\cite{Kostelecky:1995ei}
\bibitem{Kostelecky:1995ei}
V.~A.~Kosteleck\'y and M.~J.~Perry,
``Solitonic Black Holes in Gauged $N=2$ Supergravity,''
Phys.\ Lett.\ B {\bf 371} (1996) 191
[arXiv:hep-th/9512222].
%%CITATION = HEP-TH 9512222;%%

%\cite{Caldarelli:1998hg}
\bibitem{Caldarelli:1998hg}
M.~M.~Caldarelli and D.~Klemm,
``Supersymmetry of anti-de~Sitter black holes,''
Nucl.\ Phys.\ B {\bf 545} (1999) 434
[arXiv:hep-th/9808097].
%%CITATION = HEP-TH 9808097;%%

%\cite{Cacciatori:1999rp}
\bibitem{Cacciatori:1999rp}
S.~Cacciatori, D.~Klemm and D.~Zanon,
``w$_{\infty}$ algebras, conformal mechanics, and black holes,''
Class.\ Quant.\ Grav.\  {\bf 17} (2000) 1731
[arXiv:hep-th/9910065].
%%CITATION = HEP-TH 9910065;%%

%\cite{Alonso-Alberca:2000cs}
\bibitem{Alonso-Alberca:2000cs}
N.~Alonso-Alberca, P.~Meessen and T.~Ort\'{\i}n,
``Supersymmetry of topological Kerr-Newman-Taub-NUT-AdS spacetimes,''
Class.\ Quant.\ Grav.\  {\bf 17} (2000) 2783
[arXiv:hep-th/0003071].
%%CITATION = HEP-TH 0003071;%%

%\cite{Brecher:2000pa}
\bibitem{Brecher:2000pa}
D.~Brecher, A.~Chamblin and H.~S.~Reall,
``AdS/CFT in the infinite momentum frame,''
Nucl.\ Phys.\ B {\bf 607} (2001) 155
[arXiv:hep-th/0012076].
%%CITATION = HEP-TH 0012076;%%

%\cite{Plebanski:gy}
\bibitem{Plebanski:gy}
J.~F.~Pleba\'nski and M.~Demia\'nski,
``Rotating, Charged, And Uniformly Accelerating Mass In General Relativity,''
Annals Phys.\  {\bf 98} (1976) 98.
%%CITATION = APNYA,98,98;%%

%\cite{Freedman:1976aw}
\bibitem{Freedman:1976aw}
D.~Z.~Freedman and A.~Das,
``Gauge Internal Symmetry In Extended Supergravity,''
Nucl.\ Phys.\ B {\bf 120} (1977) 221.
%%CITATION = NUPHA,B120,221;%%

%\cite{Fradkin:1976xz}
\bibitem{Fradkin:1976xz}
E.~S.~Fradkin and M.~A.~Vasiliev,
``Model Of Supergravity With Minimal Electromagnetic Interaction,''
LEBEDEV-76-197
%\href{http://www.slac.stanford.edu/spires/find/hep/www?r=lebedev-76-197}{SPIRES entry}

%\cite{Siklos:1985}
\bibitem{Siklos:1985}
S.~T.~C.~Siklos, ``Lobatchevski plane gravitational waves,''
in: {\it Galaxies, axisymmetric systems and relativity},
ed.~M.~A.~H.~MacCallum, Cambridge University Press, Cambridge (1985).

%\cite{Banados:1999tw}
\bibitem{Banados:1999tw}
M.~Ba\~{n}ados, A.~Chamblin and G.~W.~Gibbons,
``Branes, AdS gravitons and Virasoro symmetry,''
Phys.\ Rev.\ D {\bf 61} (2000) 081901
[arXiv:hep-th/9911101].
%%CITATION = HEP-TH 9911101;%%

%\cite{Breitenlohner:jf}
\bibitem{Breitenlohner:jf}
P.~Breitenlohner and D.~Z.~Freedman,
``Stability in gauged extended supergravity,''
Annals Phys.\  {\bf 144} (1982) 249.
%%CITATION = APNYA,144,249;%%

\bibitem{Podolsky:1997ik}
J.~Podolsky,
``Interpretation of the Siklos solutions as exact gravitational waves in the
anti-de Sitter universe,''
Class.\ Quant.\ Grav.\  {\bf 15} (1998) 719
[arXiv:gr-qc/9801052].
%%CITATION = GR-QC 9801052;%%

%\cite{Boyer:mm}
\bibitem{Boyer:mm}
C.~P.~Boyer and J.~D.~Finley,
``Killing vectors in selfdual, euclidean Einstein spaces,''
J.\ Math.\ Phys.\  {\bf 23} (1982) 1126.
%%CITATION = JMAPA,23,1126;%%

%\cite{Kerimo:2004gz}
\bibitem{Kerimo:2004gz}
J.~Kerimo and H.~Lu,
``pp-waves in AdS gauged supergravities and supernumerary supersymmetry,''
arXiv:hep-th/0408143.
%%CITATION = HEP-TH 0408143;%%

%\cite{Seiberg:1990eb}
\bibitem{Seiberg:1990eb}
N.~Seiberg,
``Notes on quantum Liouville theory and quantum gravity,''
Prog.\ Theor.\ Phys.\ Suppl.\  {\bf 102}, 319 (1990).
%%CITATION = PTPSA,102,319;%%

%\cite{Chamseddine:1999xk}
\bibitem{Chamseddine:1999xk}
A.~H.~Chamseddine and W.~A.~Sabra,
``Magnetic strings in five-dimensional gauged supergravity theories,''
Phys.\ Lett.\ B {\bf 477} (2000) 329
[arXiv:hep-th/9911195].
%%CITATION = HEP-TH 9911195;%%

%\cite{Klemm:2000nj}
\bibitem{Klemm:2000nj}
D.~Klemm and W.~A.~Sabra,
``Supersymmetry of black strings in D = 5 gauged supergravities,''
Phys.\ Rev.\ D {\bf 62} (2000) 024003
[arXiv:hep-th/0001131].
%%CITATION = HEP-TH 0001131;%%

%\cite{Guralnik:2003we}
\bibitem{Guralnik:2003we}
G.~Guralnik, A.~Iorio, R.~Jackiw and S.~Y.~Pi,
``Dimensionally reduced gravitational Chern-Simons term and its kink,''
Annals Phys.\  {\bf 308} (2003) 222
[arXiv:hep-th/0305117].
%%CITATION = HEP-TH 0305117;%%

%\cite{Ozsvath}
\bibitem{Ozsvath}
I.~Ozsv\'ath,
``Homogeneous solutions of the Einstein-Maxwell Equations,''
J.\ Math.\ Phys.\  {\bf 6} (1965) 1255.

%\cite{Grumiller:2003ad}
\bibitem{Grumiller:2003ad}
D.~Grumiller and W.~Kummer,
``The classical solutions of the dimensionally reduced gravitational
Chern-Simons theory,''
Annals Phys.\  {\bf 308} (2003) 211
[arXiv:hep-th/0306036].
%%CITATION = HEP-TH 0306036;%%

%\cite{Bergamin:2004me}
\bibitem{Bergamin:2004me}
L.~Bergamin, D.~Grumiller, A.~Iorio and C.~Nunez,
``Chemistry of Chern-Simons supergravity: Reduction to a BPS kink, oxidation
to M-theory and thermodynamical aspects,''
arXiv:hep-th/0409273.
%%CITATION = HEP-TH 0409273;%%

%\cite{Podolsky:2002sy}
\bibitem{Podolsky:2002sy}
J.~Podolsky and M.~Ortaggio,
``Explicit Kundt type II and N solutions as gravitational waves in various
type D and O universes,''
Class.\ Quant.\ Grav.\  {\bf 20} (2003) 1685
[arXiv:gr-qc/0212073].
%%CITATION = GR-QC 0212073;%%

%\cite{Godel:ga}
\bibitem{Godel:ga}
K.~G\"odel,
``An Example Of A New Type Of Cosmological Solutions Of Einstein's Field Equations Of Gravitation,''
Rev.\ Mod.\ Phys.\  {\bf 21} (1949) 447.
%%CITATION = RMPHA,21,447;%%

%\cite{Reboucas:hn}
\bibitem{Reboucas:hn}
M.~J.~Rebou\c{c}as and J.~Tiomno,
``On The Homogeneity Of Riemannian Space-Times Of Godel Type,''
Phys.\ Rev.\ D {\bf 28} (1983) 1251.
%%CITATION = PHRVA,D28,1251;%%

%\cite{Deser:1981wh}
\bibitem{Deser:1981wh}
S.~Deser, R.~Jackiw and S.~Templeton,
``Topologically Massive Gauge Theories,''
Annals Phys.\  {\bf 140} (1982) 372
[Erratum-ibid.\  {\bf 185} (1988\ APNYA,281,409-449.2000) 406.1988\ APNYA,281,409].
%%CITATION = APNYA,140,372;%%

%\cite{Sabra:1999ux}
\bibitem{Sabra:1999ux}
W.~A.~Sabra,
``Anti-de Sitter BPS black holes in $N = 2$ gauged supergravity,''
Phys.\ Lett.\ B {\bf 458} (1999) 36
[arXiv:hep-th/9903143].
%%CITATION = HEP-TH 9903143;%%

%\cite{Chamseddine:2000bk}
\bibitem{Chamseddine:2000bk}
A.~H.~Chamseddine and W.~A.~Sabra,
``Magnetic and dyonic black holes in $D = 4$ gauged supergravity,''
Phys.\ Lett.\ B {\bf 485} (2000) 301
[arXiv:hep-th/0003213].
%%CITATION = HEP-TH 0003213;%%



\end{thebibliography}
\end{document}